\documentstyle{article}

\newcounter{eqnletter}[equation]
\catcode`\@=11
\@addtoreset{equation}{section}
\catcode`\@=12

\begin{document}
 \begin{centering}
{ \Large \bf Phase transitions and random matrices}\\

\vskip 1cm  G.M.Cicuta  \\

 Dept.of Physics, Univ.of
Parma, \\ Viale delle Scienze,
 43100 Parma , Italy.\\
E-mail : cicuta@fis.unipr.it
 \vskip 1cm

{\bf Abstract  }\\
\end{centering}
Phase transitions generically occur in random matrix models as the
parameters in the joint probability distribution of the random
variables are varied. They affect all main features of the theory
and the interpretation of statistical models . In this paper a
brief review of phase transitions in invariant ensembles is
provided, with some comments to the singular values decomposition
in complex non-hermitian ensembles.
\\ \vskip 2cm

\section  {Phase transitions in invariant hermitian ensembles. }

Random matrix ensembles have been extensively studied for several decades,
since the early works of E.Wigner and F.Dyson, as effective mathematical reference models
for the descriptions of statistical properties of the spectra of complex
physical systems. In the past twenty years new  applications spurned
a large literature both in theoretical physics and among mathematicians.
Several monographs review different sides of the physics literature
of the past few decades,
such as  \cite{Amb} ,  \cite{Bee}, \cite {brw} \cite{dem}  \cite{DiF}, \cite{Guh}
 \cite{ros} \cite{ss} . Their
 combined bibliography, although very incomplete, exceeds a thousand papers.
Sets of lecture notes are \cite{tra} \cite{gro} \cite{amb}\cite{da}  
\cite{mul} \cite {pol5} . The classic reference is  Mehta's book
\cite{Meh}.\\

For a long time studies and applications of random matrix theory in large part were limited to the choice of
 gaussian random variables for the independent entries of the random matrix. This was due both to the dominant
  role of the normal distribution in probability theory as well as to the nice analytic results which were obtained.
   Increasingly, in the past two decades, a wide variety of matrix ensembles were considered, where the joint
    probability distribution for the random entries depends on a number of parameters.
    As the latter are allowed to change, the generic occurrence of phase transitions  emerged.\\
In this section, I begin by recalling
 the problem in the easiest case, the
invariant ensemble  of hermitian matrices .\\

Let $ H=(H_{ij})_{i,j=1,..,N}$
 be hermitian random matrix with joint probability density for the independent entries
\begin{eqnarray}
P \left(H_{11},H_{12},..,H_{NN}\right)\, dH &=&e^{-N {\rm Tr} \,
V(H) }\; dH  \Bigg/ \int e^{-N {\rm Tr} \, V(H) }\; dH \quad ,
\nonumber \\ V(H)&=&\frac{1}{2}a_2 H^2+\frac{1}{4}a_4 H^4+..
+\frac{1}{2p}a_{2p}H^{2p} \quad, \quad a_{2p}>0
 \quad,
\nonumber \\
 \qquad
 dH &=& \prod_{i<j} (d \, Re\, H_{ij}\, d\, Im \, H_{ij})\; \prod_i d\, H_{ii}
\label{a.1}
\end{eqnarray}

We are interested in the partition function $Z_N(a_2, a_4,..,a_{2p})$,
the free energy $F_N(a_2, a_4,..,a_{2p})=- \log Z_N(a_2, a_4,..,a_{2p})
 $, the "one point resolvent" $G_N(z) \,$ , the connected correlator
$G_N^{(c)} (z_1, z_2)$
\begin{eqnarray}
&&  Z_N(a_2, a_4,..,a_{2p}) = \int e^{-N {\rm Tr} \, V(H) }\; dH
\quad , \quad
G_N(z)=\frac{1}{N} {\rm Tr} < \frac{1}{z-H}>
\nonumber\\
&& G_N^{(c)} (z_1, z_2)=<{\rm Tr} \frac{1}{z_1-H} {\rm Tr} \frac{1}{z_2-H}  >
-<{\rm Tr} \frac{1}{z_1-H}> <{\rm Tr} \frac{1}{z_2-H}  >
\nonumber \\
\label{a.2}
\end{eqnarray}
The name invariant ensemble for the ensemble of these hermitian matrices
reminds that since the density $P \left(H_{11},H_{12},..,H_{NN}\right)$ is
invariant under a similarity transformation $H \to UHU^{-1}$ with arbitrary
unitary matrix $U$, most of the interesting quantities, like those in eqs.(\ref{a.2})
may be evaluated from the joint probability density of the eigenvalues.\\

Also  important are the monic polynomials $P_n(z)=z^n+0(z^{n-1})$,
orthogonal on the real line with the weight $e^{-N \, V(z)}$
\begin{eqnarray}
 \int_{-\infty}^{\infty} P_n(z) P_m(z) \, e^{-N \, V(z)} dz &=& h_n \, \delta_{nm}
\quad , \nonumber \\
 z\, P_n(z)=  P_{n+1}(z)+R_n \, P_{n-1}(z) \quad &,& \quad
R_n=\frac{h_n}{h_{n-1}} >0  \quad,
 \nonumber \\
Z_N(a_2, a_4,..,a_{2p})&=&N! \,(h_0)^N \,\prod_{n=1}^{N-1} (R_n)^{N-n}
\label{a.3}
\end{eqnarray}
One  obtains  a non-linear recursion relation for the coefficients $R_n$. For instance
if $V(x)=\frac{a_2}{2} x^2+\frac{a_4}{4} x^4+\frac{a_6}{6} x^6$ one has the recurrence
 relation (see \cite{Le1} ,  \cite{ IK} ,  \cite {CGM} )
\begin{eqnarray}
\frac{n}{N}=R_n \!\!& \!\!\bigg[ \!\!& \! \!  a_2 + a_4
\left(R_{n-1}+R_n+R_{n+1} \right)+a_6 \left( R_{n-1}+R_n+R_{n+1}
\right)^2 + \nonumber \\ &+&  a_6
\left(R_{n-2}R_{n-1}-R_{n-1}R_{n+1}+R_{n+1}R_{n+2} \right) \bigg]
\label{a.4}
\end{eqnarray}
occasionally called "pre-string equation" or the Freud equation.
One also introduces the  set of orthonormal functions $\psi_n(z)$
and the two point kernel $K_N(x,y)$ in terms of which all
$n-$point correlation functions are expressible
\begin{eqnarray}
\psi_n(z)=\frac{1}{\sqrt{h_n}} e^{-N V(z)/2} \,P_n(z) \quad, \quad
K_N(x,y) =\sum_{j=0}^{N-1} \psi_j(x) \psi_j(y) \nonumber\\
\label{a.5}
\end{eqnarray}
If all the coefficients $a_{2k}>0$ the statistics of the
eigenvalues may be evaluated in the limit  $N \to \infty$ , see
\cite{bou} ,
  $\lim_{N \to \infty}G_N(z) =G(z)$ is holomorphic in the complex z
plane, except for a segment  $(-A,A)$ on the real axis. Furthermore
\begin{eqnarray}
 G(z)=\int _{-A}^A d\mu \, \frac{ \rho(\mu)}{z-\mu} \quad , \quad
\rho(\mu)=\lim_{N \to \infty}
\frac{1}{N} {\rm Tr} < \delta( \mu-H)>=\lim_{N \to \infty} K_N(\mu, \mu)
\nonumber
\end{eqnarray}
The limiting density of eigenvalues $\rho(\lambda)$
 , the one point correlation function, is the unique solution
of the integral equation :
\begin{eqnarray}
V'(\lambda) = 2 \, {\cal P} \int_{-A}^A d\mu \, \frac{
\rho(\mu)}{\lambda-\mu} \quad  , \quad \int_{-A}^A  \rho(\lambda)
\,d\lambda =1
\label{a.6}
\end{eqnarray}
$\rho(\lambda)>0$ on its support $(-A,A)$ , it vanishes as a square root
at the boundary $\rho(\lambda) \sim (A-|\lambda |)^{1/2} $. Furthermore
the coefficients $R_n$  approach a smooth  limit $R(\frac{n}{N}) \sim
R(x)$ and
\begin{eqnarray}
F(a_2, a_4,..,a_{2p})\!\!\!&=&\!\!\!N^2 \Bigg[ \int_{-A}^A d \lambda \, \rho (\lambda)\,
 V(\lambda) -\int \!\!\int_{-A}^A d \lambda \,d \mu
 \, \rho (\lambda)\,  \rho (\mu)\, \log|\lambda-\mu|
+O(\frac{1}{N^2}) \Bigg]=
\nonumber \\
&=&N^2 \Bigg[-\int_0^1 dx \,(1-x)\, \log \,R(x)
-\frac{1}{N} \log \, h_0
+O(\frac{1}{N^2}) \Bigg]
\nonumber
\end{eqnarray}
The free energy $F(a_2, a_4,..,a_{2p})$ is analytic in the couplings $(a_2, a_4,..,a_{2p})$.
The saddle point solution is equivalent to the resummation of the planar graphs,
it is equivalent to the solution from the recurrence relations for $R_n$
and to other techniques, such as the loop equations. The orthogonal
polynomial technique and  the loop equations are superior to evaluate
in a systematic way the terms in the series in the parameter $\frac{1}{(N^2)^k}$
corresponding to the resummation of the graphs which are embeddeble
 on orientable surfaces with $k$ handles \cite{tHo} .

It was also proved that the connected two point correlators exhibit two different
forms of universality (that is independence from the set of coefficients $\{ a_{2k}
\}$ but depend only on the endpoints $\pm A) : $

Global universality. The limiting connected density-density correlator
$\rho_c(\lambda , \lambda') \, ,$  after smoothing over a scale much larger
than the level spacing $\Delta_N$ , is \cite{ajm} \cite {bre1}
\begin{eqnarray}
\rho_c(\lambda , \lambda') =-\frac{1}{2 \pi^2(\lambda -
\lambda')^2} \frac{A^2-\lambda \, \lambda'}{\sqrt{A^2-\lambda^2}
\sqrt{A^2-\lambda'^2}} \quad , \quad \lambda \neq \lambda'
\nonumber
\end{eqnarray}
Local universality. The limiting two point kernel $K(x,y)$ has the sine law
\cite {pss}
for eigenvalues in the bulk of the spectrum, measured in units of $\Delta_N$
\begin{eqnarray}
K_{ {\rm bulk}}(s, s')= \frac{\sin[ \pi (s-s')]}{\pi  (s-s')} \quad, \quad
s=\lambda/ \Delta_N \; , \; s'=\lambda'/\Delta_N
\nonumber
\end{eqnarray}
or the Airy law , close to the tail of the spectrum (the soft edge) \cite {bow}
\cite {baw}
\begin{eqnarray}
K_{ {\rm soft}}(s, s')= \frac{Ai(s)\, Ai'(s')-Ai(s')\,Ai'(s)}{s-s'}
\quad , \quad s \sim N^{2/3} \left(\frac{\lambda}{A}-1 \right)
\nonumber
\end{eqnarray}
A general derivation of spectral correlators is given in the recent paper by
Kanzieper and Freilikher \cite{kf}. By following Shohat  \cite{sh}, the
recurrence relation for the orthogonal polynomials $P_n(x)$ is turned into an
exact second-order differential equation for the orthogonal functions $\psi_n(z)$
\begin{eqnarray}
\frac{d^2 \psi_n(\lambda)}{d\lambda^2}- {\cal F}_n(\lambda)
\frac{d \psi_n(\lambda)}{d\lambda}+{\cal G}_n(\lambda)\psi_n(\lambda)=0
\nonumber
\end{eqnarray}
Because of the smooth behavior of the coefficients $R_n$ , in the
case where the eigenvalues have  support on a single segment, the
complicated forms $ {\cal F}_n(\lambda)$ , ${\cal G}_n(\lambda)$
simplify at large order and the differential equation leads to the
global and the local universality results mentioned above.
\footnote{If a logarithmic singularity is present at the origin,
the Bessel law is derived
\begin{eqnarray}
K_{ {\rm origin}}(s, s')= \frac{\pi}{2} (s\, s')^{1/2} \frac{J_{\alpha+1/2}(\pi s)
J_{\alpha-1/2}(\pi s')-J_{\alpha-1/2}(\pi s)J_{\alpha+1/2}(\pi s')}{s-s'}
\nonumber
\end{eqnarray}
where $s$ and $s'$ are scaled by the level spacing $\Delta_N(0)$
near the spectrum origin, $s=\lambda/\Delta_N(0)$ , and
$\alpha>-1/2$.  For the simpler situations where the orthogonal
polynomials are classical the Bessel law had been derived in \cite
{br}  \cite {na1}  \cite {fo}  \cite {na2} .}\\

All the above is deeply affected by {\bf phase transitions.} \\

The limiting eigenvalue density $\rho(\lambda ; a_2,
a_4,..,a_{2p})$  continued to negative values for one or several
coefficients $a_{2k}$ (while keeping $a_{2p}>0$ ) may not be
positive definite. Then the integral equation (\ref{a.6}) allows
new solutions with eigenvalue density positive definite with
support on two or more segments of the real axis. The correct
solution minimizes the free energy. As one explores the space of
the parameters, the  free energy is evaluated on different saddle
point solutions, which may coincide for certain critical values of
the parameters. Then the  free energy is usually a continuous but
not analytic function of the parameters.\\ The recursion
coefficients $R_n$ no longer have a smooth "continuous" limit,
which makes more difficult the orthogonal polynomials solution. \\
The lack of analyticity of the free energy at critical values of
the parameters is analogous to phase transitions related to
spontaneous symmetry breaking in classical statistical mechanics.
For instance, in the simplest case of potential
$V(x)=\frac{a_2}{2} x^2+\frac{a_4}{4} x^4$ , first analysed for
negative values of $a_2$ in papers \cite {shi} \cite {cmm}, it is
convenient to add a linear term , which explicitly breaks the
$Z_2$ symmetry, $V(x)=a_1 x+ \frac{a_2}{2} x^2+\frac{a_4}{4} x^4$
, next perform the "thermodynamic limit" $N \to \infty$, finally
remove the symmetry breaking term $a_1 \to 0$, see \cite{ilg} ,
\cite {cmm2} . This allows the evaluation of the "order parameter"
$<{\rm Tr} H>$
\begin{eqnarray}
\lim_{a_1 \to 0} \lim_{N \to \infty} <{\rm Tr} H>= {\rm sign}(a_1) \, \theta(
-a_2-2 \sqrt{a_4}) f(a_2, a_4)
\nonumber
\end{eqnarray}
For the simplest case $V(x)=\frac{a_2}{2} x^2+\frac{a_4}{4} x^4$,
it was shown that the correct ansatz for the  recursion
coefficients $R_n$ , if $a_2<-2 \sqrt{a_4}$ is that the even
$R_{2n}$ and odd $R_{2n+1}$ approach two different smooth
"continuous" functions \cite{mol}. This period-two ansatz requires
a little generalization in the
 case which includes the infinitesimal symmetry breaking term  \cite{mol2}
 \cite{bro} because it leads to recurrence relations
\begin{eqnarray}
 z\, P_n(z)=  P_{n+1}(z)+ S_n\, P_n(z)+ R_n \, P_{n-1}(z) \quad &,& \quad
R_n=\frac{h_n}{h_{n-1}} >0  \quad,
\nonumber
\end{eqnarray}
and it explains the origin of the period-two ansatz. \\

However in the next simplest case , like
 $V(x)=\frac{a_2}{2} x^2+\frac{a_4}{4} x^4+\frac{a_6}{6} x^6$ , or higher
order polynomials, corresponding to multiple well potentials, which may not be
degenerate,  the
behaviour of the coefficients $R_n$ is erratic and it is difficult
to reproduce the results of the saddle point analysis by the orthogonal
polynomials \cite {ju2} \cite{le1}  \cite{le2}  \cite{sas} \cite{sen}. \\
The lines of "phase transition" in the parameter space, related to the
continuation to negative values
of the coefficient $a_{2p}$ of  the monomial of highest order may be found
in terms of previously discussed phase transitions by the addition of an
infinitesimal  monomial $\epsilon \, x^{2p+2}$, see for instance
\cite{cmm3} \cite {ju1}.\\

Connected correlators, when the support of the eigenvalue density is
two segments, were shown \cite{aa}  \cite {ake} to have a different form of global
universality, involving elliptic integrals. In the case of multicritical behaviour
the local universality form has a modified Bessel law \cite{admn}.\\

Important recent works seem to be so powerful and comprehensive to
solve the above mentioned ambiguities. The Freud equation is
expressed in a matrix Lax representation and the semiclassical
asymptotics of the functions $\psi_n(z)$ is obtained in the whole
complex $z$ plane by solving a matrix Riemann-Hilbert problem,
following earlier works \cite {fi1} \cite {fi2}. Very useful is
the non-linear steepest descent method devised by Deift and Zhou
\cite{dz1} \cite{dz2}. This set of works \cite{bl} \cite{dz3} \cite{dk1}
\cite{dk2} not only provide rigorous and more general solutions for
methods and ansatzes previously used, but it seem to provide
answers also for the models previously left unsolved (like the
asymptotics of recurrence coefficients $R_n$ in general cases).
The recent and difficult developments will require some time to be
exploited by physicists.\\

Finally, while almost all investigations related to the invariant ensemble
of random matrices considered a probability distribution of the form of
exponential of a polynomial like in eq.(\ref{a.1}), with possible addition
of logarithmic terms, like the Penner model and the Kontsevich model,
there exist  probability distributions, still invariant
under diagonalization by unitary matrices, which lead to different
forms for the connected correlators \cite{ci6}. Then the classification of universality
classes perhaps is not yet complete, even in invariant one-matrix
hermitian ensembles.

\section {Further matrix ensembles and singular value decomposition  in complex
non-hermitian random matrices}

Random matrix models more general than the hermitian one-matrix invariant
ensemble are often more interesting because of the possibility to describe
more interesting statistical models. The analytic solution of multi-matrix
 models both in the "perturbative phase" or in different phases is more
complex. While an ensemble of hermitian random matrices describes
triangulations of random orientable surfaces, multi-matrix
ensembles are suitable to describe models of classical statistical
mechanics on a random two-dimensional lattice. After the
breakthrough of the Ising model \cite {ka1} , \cite {ka2}, it was
possible to study random walks and loops \cite {dup} , $O(N)$
model \cite{gau} \cite{ko} \cite{kos} \cite{ey1} \cite{ey2} \cite
{dur} , the Potts model \cite {ka8} \cite {da2} \cite {pz},
 surfaces with holes  \cite{ka3} ,
 \cite {ci5}, a special case of $8$-vertex model  \cite{ka4} \cite {pz2} ,
the chiral random matrix which simulate the spontaneously broken phase
transition of QCD \cite{ve1} \cite{no1} \cite{shu} \cite{sim}  \cite{pol2} \cite{ber}.
 Often it was possible to evaluate
 the   critical exponents at phase transition. Multi-matrix models of hermitian
 matrices are also a good framework for combinatorial problems
like the four-color theorem \cite{cix} \cite{ck} \cite {ek} or the enumeration
of meanders \cite{di2} \cite{mak}.
 Most influential, for quantum field theorists were the phase
transitions in models of random unitary matrices, describing
one-plaquette of the lattice formulation of QCD \cite{gw} \cite
{bg} \cite{jur} , the saddle-point solution of the one-matrix
ensemble \cite{bip} , the Witten conjecture of a master field
\cite{wit}.\\

In this section I shall recall the singular value decomposition, which
plays a role in the analysis of models with rectangular random matrices
and square complex non-hermitian matrices.\\

 Let $ \phi=(\phi_{ij})_{i=1,..,N \atop j=1,..,M}$
 be rectangular random matrix with entries $\phi_{ij}$ real or complex numbers,
and joint probability distribution $P(\phi)$ invariant under  $\phi \to U \phi V$
, with $U$ unitary of  order $N$ , $V$ unitary of  order $M$
\begin{eqnarray}
&& P(\phi) \,d\phi \,=\, e^{-N {\rm Tr} \, V(\phi^{\dag} \phi) }\; d\phi  \Bigg/ \int
e^{-N {\rm Tr} \,
 V(\phi^{\dag} \phi) }\; d\phi
\nonumber \\ && \; V(\phi^{\dag} \phi)=\frac{1}{2}a_2 (\phi^{\dag}
\phi)+\frac{1}{4}a_4 (\phi^{\dag}\phi)^2+..
+\frac{1}{2p}a_{2p}(\phi^{\dag} \phi)^{p} \quad, \quad a_{2p}>0
 \quad,
\nonumber \\
&& d\phi=\!\prod_{i=1,..,N \atop j=1,..,M} d \, \phi_{ij} \; , \quad
{\rm if \; \phi \; is \; real}; \quad {\rm or}
\nonumber\\
&& d\phi=\!\prod_{i=1,..,N \atop j=1,..,M} (d \, Re\,
\phi_{ij}\, d\, Im \, \phi_{ij})\; ,
 \quad {\rm if \; \phi \; is \; complex}
 \label{b.1}
\end{eqnarray}
The hermitian matrices $\phi^{\dag} \phi$ and $\phi \, \phi^{\dag}
$ are positive semi-definite, have the same  non-vanishing
eigenvalues $t_i=\sigma_i^2$ , where $ \sigma_i$ are the singular
values of $\phi$ , themselves positive definite. It is
straightforward to evaluate the statistics of the singular values
in the limit $N \to \infty$ , $M \to \infty$, while the ratio
$N/M=L$ is fixed, by the ordinary saddle point analysis.
 Let us consider first $L \geq1$. The probability distribution (\ref{b.1}) is
\begin{eqnarray}
e^{-N  \sum_{i=1,..,M} \, V(\sigma_i^2)} \; J(\sigma_i)
\prod_{i=1,..,M} d\sigma_i
 \Bigg/ \int e^{-N  \, \sum_{i=1,..,M} \, V(\sigma_i^2)} J(\sigma_i) \prod_{i=1,..,M}
 d\sigma_i
 \nonumber
\end{eqnarray}
where the Jacobian $J(\sigma_i)$ may be evaluated with help from
\cite{hel} ,\cite{hua}
\begin{eqnarray}
J(\sigma_i) &=&\prod_{i=1,..,M} (\sigma_i)^{N-M} \prod_{1 \leq i <
j \leq  M} |\sigma_i^2 -\sigma_j^2| \; , \qquad {\rm if \; \phi \;
is \; real}, \nonumber \\ &=&\prod_{i=1,..,M} (\sigma_i)^{2N-2M+1}
\prod_{1 \leq i < j \leq  M} |\sigma_i^2 -\sigma_j^2|^2 \; ,
\qquad {\rm if \; \phi \; is \; complex}
 \nonumber
\end{eqnarray}
For the simple case $V(\phi^{\dag} \phi)=\frac{1}{2}a_2
(\phi^{\dag} \phi)+\frac{1}{4}a_4 (\phi^{\dag} \phi)^2$ one easily
obtains \cite {ci4} for
 complex rectangular matrix $\phi$
\begin{eqnarray}
G(z)&=&\lim_{N \to \infty \atop M \to \infty} \frac{1}{M}
 {\rm Tr}_M <\frac{1}{z-\phi^{\dag} \phi}>=
\int_A^B dt \frac{u(t)}{z-t} \quad , \quad 0 \leq A \leq B
\nonumber \\ u(t)&=&\frac{1}{\pi} \sqrt{(B-t)(t-A)} \left(
\frac{a_4}{4}+\frac{a_4(A+B)}{8 t}+ \frac{a_2}{2 t} \right)
 \label{b.3}
\end{eqnarray}
The extrema $A, B$ are given by the usual couple of algebraic
equation. Then for $a_4=0$ one finds the distribution of singular
values for rectangular matrices, which is the generalization of
Wigner "semi-circle law"
\begin{eqnarray}
u(\sigma)= \frac{a_2}{\pi} \frac{\sqrt{(B-\sigma^2 )(
\sigma^2-A)}}{  \sigma} \quad &,& \quad {\rm for} \; \; \sqrt{A}
\leq \sigma \leq \sqrt{B} \nonumber \\ A= \left( \frac{
\sqrt{L}-1}{a_2} \right)^2 \quad &,& \quad B= \left( \frac{
\sqrt{L}+1}{a_2} \right)^2
 \nonumber
\end{eqnarray}
Returning now to eq.(\ref{b.3})  for square complex matrices ,
$L=1$, one finds a "perturbative" phase for $a_2>-2 \sqrt{a_4}$
with $A=0$ where observables correspond to resummation of planar
graphs, and a "non-perturbative" phase for $a_2<-2 \sqrt{a_4}$
where $A>0$ , quite similar to the random hermitian case. These
results were rediscovered by several authors \cite{vip1}
\cite{vip2} \cite{fz} who introduced the hermitian matrix $H$ and
the partition function ${\cal Z}$
\begin{eqnarray}
&& H=\left(  \begin{array}{cc} 0 & \phi \\ \phi^{\dag} & 0
\end {array}
\right) \quad , \quad {\cal Z}=\int DH \; e^{-\beta {\rm Tr}
V(H^{\dag}H)} \sim
 \nonumber \\
&\sim & \int_{-\infty}^{\infty}\left[ \prod_{i=1}^M dx_i \,
e^{-2\beta \, V(x_i^2)} \right]  \prod_{i=1}^M |x_i|^{2N-2M+1}
\prod_{1 \leq i<j  \leq M} (x_i^2-x_j^2)^2
 \nonumber
\end{eqnarray}
The eigenvalues $x_i$ of the "chiral" matrix $H$ are in two to one
correspondence with the singular values $\sigma_i$ of $A$ : $x_i=
\pm \sigma_i$. The technique of using the auxiliary matrix $H$ in
the study of square complex non-hermitian matrix $\phi$ was
developed by \cite {fz2} \cite {fz3} \cite{bbhz} into a powerful
method to obtain the distribution $\rho(x,y)$ of complex
eigenvalues $\lambda_i$ , see also the similar and simultaneous
paper \cite {pol1}  . Indeed
\begin{eqnarray}
\rho(x,y)&=&\frac{1}{N}\sum_{i=1}^N <\delta (x-{\rm
Re}\,\lambda_i) \delta (y-{\rm Im}\,\lambda_i)>= \nonumber \\ &=&
\frac{1}{\pi}\frac{\partial}{\partial \,
z}\frac{\partial}{\partial \, z^*} \frac{1}{N}<{\rm Tr}_{(N)}
\log(z-\phi)(z^*-\phi^{\dag})>= \nonumber \\ &=&
\frac{1}{\pi}\frac{\partial}{\partial \,
z}\frac{\partial}{\partial \, z^*} \frac{1}{N}\left[ <{\rm
Tr}_{(2N)} \log H> -i \pi N^2 \right]
 \nonumber
\end{eqnarray}
where now
\begin{eqnarray}
H=\left(  \begin{array}{cc} 0 & \phi-z \\ \phi^{\dag}-z^* & 0
\end {array}
\right)
 \nonumber
\end{eqnarray}
Diagrammatic rules may then be used to evaluate the resolvent
${\cal G}(\eta ; z, z^*)$
\begin{eqnarray}
{\cal G}(\eta ; z, z^*)=\frac{1}{2 N} <{\rm Tr}_{(2N)}
\frac{1}{\eta-H}>= \frac{\eta}{N}<{\rm
Tr}_{N}\frac{1}{\eta^2-(z^*-\phi^{\dag})(z-\phi)}>
 \nonumber
\end{eqnarray}
Finally in terms of the integrated density of eigenvalues of $H$ ,
$\Omega(\mu ; z, z^*)=\frac{1}{2N} <{\rm Tr}_{(2N)}
\theta(\mu-H)>$ , Feinberg and Zee \cite{fz3} obtain
\begin{eqnarray}
\rho(x,y)=-\frac{4}{\pi}\int_0^{\infty}d\mu \;
\frac{\partial}{\partial \, z}\frac{\partial}{\partial \, z^*}
\frac{\Omega(\mu ; z, z^*)}{\mu}
 \nonumber
\end{eqnarray}
As specific examples, Feinberg and Zee consider probability
distributions $P(\phi, \phi^{\dag})$ for the complex matrix $\phi$
of the form (\ref{b.1}), invariant under $\phi \to e^{i \alpha}
\phi $ , $\phi^{\dag} \to e^{-i \alpha} \phi^{\dag}$, where it is
natural to expect that the distribution of complex eigenvalues
$\rho(x,y)$ has rotational symmetry $\rho(x,y)=\rho(r)/(2\pi)$, as
it indeed happens with the Ginibre gaussian ensemble \cite{gin}.
For the simple model with only two coefficients $a_2$ , $a_4$ they
find that for $a_2>-2\sqrt{a_4}$ the complex eigenvalues fill
(non-uniformly)
 a disk centered at the origin, while for $a_2<-2\sqrt{a_4}$ they fill
a ring. This is expected from the analogous distribution of the
singular values of the matrix $\phi$ . Next they prove the
surprising "single ring theorem" asserting that for a generic
polynomial potential of the form (\ref{b.1}), even in the
multiple-well cases where the distribution of the singular values
of the matrix $\phi$ has support on several segments of the
positive real axis, the distribution of eigenvalues of the matrix
$\phi$ may only have one ring at most. It seems to me that in such
cases the assumption of rotational symmetry should be checked. If
it turns out, by using an explicitly symmetry breaking term to be
removed after the thermodynamic limit $N \to \infty$, that one
obtains different distributions, dependent on the direction of the
symmetry breaking probe, it would be a remarkable example of
spontaneous symmetry breaking of rotational symmetry.

\section {Conclusion.}

Phase transitions generically occur in the study of ensembles of
random matrices, as the parameters in the joint probability
distribution of the random variables are varied. They are
important for the physics interpretation of statistical models and
they affect all the main features of random matrix theory. In
invariant one-matrix ensembles recent progress of mathematicians
seems to solve long standing problems related to multi-cut
solutions with no symmetry. Phase transitions in ensembles of
complex non-hermitian matrices were recently explored and it is
likely that a richer variety of phase transitions will be
discovered. Random matrix ensembles with a preferential basis,
like band matrices were studied since the beginning of random
matrix theory , see for instance \cite {lm} \cite {cpv} \cite {fm} \cite {si} .
 Even the simplest cases of tridiagonal matrices
with random site or random hopping could be analytically solved
only for a very limited choice of the probability distribution .
Well known discrepancies between the moment method and numerical
methods suggest the presence of phase transitions which seem more
difficult to analyse than in case of invariant ensembles.


\begin{thebibliography}{99}


\bibitem {ake} G.Akemann, Higher genus correlators for the hermitian matrix
model with multiple cuts,  Nucl.Phys. {\bf B482} (1996) 403 ; the generalization
for singular values of complex matrices is G.Akemann, Universal correlators
for multi-arc complex matrix models,  Nucl.Phys. {\bf B507} (1997) 475.

\bibitem {ci6} G.Akemann, G.M.Cicuta, L.Molinari, G.Vernizzi, Compact support
probability distributions in random matrix theory, Phys.Rev.{\bf
E59} (1999) 1489 , and Non-universality of compact support
probability distributions in random matrix theory,
 Phys.Rev.{\bf E60} (1999) 5287-5292.

\bibitem {admn} G.Akemann, P.H.Damgaard, U.Magnea, S.M.Nishigaki, Multicritical
microscopic spectral correlators of hermitian and complex matrices, Nucl.Phys.
 {\bf B519} (1998) 682.
\bibitem {ajm} J.Ambjorn, J.Jurkiewicz, Y.Makeenko, Multiloop
correlators for two-dimensional quantum gravity, Phys.Lett. {\bf B251} (1990)
 517 ; the evaluation of correlators in $1/N^2$ expansion was also shown to be
universal in J.Ambjorn, L.Chekhov, C.F.Kristjansen, Y.Makeenko, Matrix model
calculations beyond the spherical limit, Nucl.Phys. {\bf B404} (1993) 127
and erratum in Nucl.Phys. {\bf B449} (1995) 681 ; and generalized to
supermatrices in \cite{ple}.
\bibitem {amb} J.Ambjorn, Quantization of geometry, in 1994 Les Houches,
ed. F.David, P.Ginsparg, J.Zinn-Justin, North-Holland 1996.
\bibitem {aa} J.Ambjorn, G.Akemann, New universal spectral correlators,
J.Phys. {\bf A29} (1996) L555.
\bibitem {Amb} J.Ambjorn, B.Durhuus, T.Jonsson, Quantum Geometry,
Cambridge Univ.Press 1997.
\bibitem {vip1} A.Anderson, R.C.Myers, V.Periwal, Branched polymers from a
double-scaling limit of matrix models, Nucl.Phys. {\bf B360} (1991) 463-479.
\bibitem {baw} E.L.Basor, H.Widom, Determinants of Airy operators and applications
 to random matrices, J.Stat.Phys. {\bf 96} (1999) 1-20.
\bibitem {Bee}  C.W.J.Beenakker,  Random-matrix theory of
 quantum transport,
Rev.Mod. Phys. {\bf 69} (1997) 731.
\bibitem {ber} Berbenni-Bitsch, M.E.Meyer, S.Schafer, J.J.Verbaarschot, T.Wettig,
Microscopic universality in the spectrum of the lattice Dirac operator,
Phys.Rev.Lett. {\bf 80} (1998) 1146.
\bibitem {bl} P.Bleher, A.Its, Semiclassical asymptotics of orthogonal
polynomials, Riemann-Hilbert problem, and universality in the matrix model,
(1997) pag.1-105.
\bibitem {ka2} D.V.Boulatov, V.A.Kazakov, The Ising model on a random planar lattice : the structure of phase transition and the exact critical exponents, 
Phys.Lett.{\bf B186} (1987) 379.
\bibitem {bou} A.Boutet de Monvel, L.Pastur, M.Shcherbina, On the statistical
mechanics approach in the random matrix theory : integrated density of states,
J.Stat.Phys. {\bf 79} (1995) 585.
\bibitem {bow} M.J.Bowick, E.Brezin, Universal scaling of the tail of the density of eigenvalues
 in random matrix models, Phys.Lett. {\bf B268} (1991) 21.
\bibitem {bip} E.Brezin, C.Itzykson, G.Parisi, J.B.Zuber , Planar diagrams, Comm.
Math.Phys. {\bf 59} (1978) 35.
\bibitem {bg} E.Brezin, D.J.Gross, The external field problem in the large $N$ limit
of QCD, Phys.Lett. {\bf B97} (1980) 120.
\bibitem {brw} E.Brezin, S.R.Wadia, The large $N$ expansion in quantum field
theory and statistical physics, World scientific 1993.
\bibitem {brz} E.Brezin, A.Zee , Universality of the correlations between
eigenvalues of large random matrices ,  Nucl.Phys. {\bf B402} (1993) 613.
\bibitem {bre1} E.Brezin, A.Zee , Correlation functions in disordered systems,
Phys.Rev {\bf E 49} (1994) 2588.
\bibitem {bbhz}  E.Brezin, S.Hikami , A.Zee , Oscillating density of states near zero
energy for matrices made of blocks with possible application to the random flux
problem, Nucl.Phys. {\bf B464} (1996) 411.
\bibitem {br} B.V.Bronk, Exponential ensemble for random matrices, J.Math.Phys.
{\bf 6} (1965) 228.
\bibitem {bro} R.C.Brower, N.Deo, S.Jain, C-I Tan, Symmetry breaking in the
double-well hermitian matrix models, Nucl.Phys. {\bf B405} (1993) 166-187.
\bibitem {ck} L.Ckekhov, C.Kristjansen , Hermitian matrix model with plaquette
interaction, Nucl.Phys. {\bf B479} (1996) 683.
\bibitem {cmm} G.M.Cicuta, L.Molinari, E.Montaldi, Large N phase transitions in low 
dimensions, Mod.Phys.Lett.{\bf1} (1986) 125-129.
\bibitem {cmm2}  G.M.Cicuta, L.Molinari, E.Montaldi, Large-N spontaneous
magnetization in zero dimension, J.Phys.{\bf A 20} (1987) L67-L70.
\bibitem {cmm3}  G.M.Cicuta, L.Molinari, E.Montaldi, Multicritical points
in matrix models, J.Phys.{\bf A 23} (1990) L421-L425.
\bibitem {cix} G.M.Cicuta, L.Molinari, E.Montaldi,  Matrix models and graph
colouring, Phys.Lett.{\bf B306} (1993) 245.
\bibitem {ci4} G.M.Cicuta, L.Molinari, E.Montaldi, R.Riva , Large rectangular
random matrices, J.Math.Phys. {\bf 28} (1987) 1716. This work follows
the previous investigations A.Barbieri, G.M.Cicuta, E.Montaldi, Nuovo Cimento
{\bf A84} (1984) 173 and C.M.Canali,  G.M.Cicuta, L.Molinari, E.Montaldi,
Nucl.Phys. {\bf B265} (1986) 485.
\bibitem {ci5} G.M.Cicuta, L.Molinari, E.Montaldi, S.Stramaglia, A matrix model
for random surfaces with dynamical holes, J. Phys. {\bf A29} (1996) 3769-3785.
\bibitem {cpv} A.Crisanti, G.Paladin, A.Vulpiani, Products of random matrices in 
statistical physics , Springer series in solid-state sciences 104, Springer-Verlag 1993.
\bibitem {CGM}  C.Crnkovic, P.Ginsparg, G.Moore, The Ising model,
the Yang-Lee edge
singularity and the 2$D$ quantum gravity, Phys.Lett.{\bf 237B} (1990) 196-201.
\bibitem {da2} J.M.Daul , Q-states Potts model on a random planar lattice,
hep-th/9502014.
\bibitem {da} F.David, Simplicial quantum gravity and random lattices, in
Gravitation and Quantization, Les Houches 1992 Session LVII, B.Julia and J.Zinn-Justin eds., North Holland 1995.
\bibitem {dz1} P.Deift and X.Zhou, A steepest descent method for
oscillatory Riemann-Hilbert problems. Asymptotics for the mKdV
equation, Ann. of Math. {\bf 137} (1993) 295-370.
\bibitem {dz2} P.Deift and X.Zhou, Asymptotics for the Painleve II
equation, Comm.Pure and Appl.Math. {\bf 48} (1995) 277-337.
\bibitem {dz3} P.Deift , A.R.Its, X.Zhou, A Riemann-Hilbert approach to
asymptotic problems arising in the theory of random matrix models
and also in the theory of integrable statistical mechanics, Ann.of Math.
{\bf 146} (1997) 149-235.
\bibitem {dk1} P.Deift, T.Kriecherbauer, K.T-R McLaughlin,
S.Venakides, X.Zhou, Strong asymptotics of orthogonal polynomials
with respect to exponential weights, preprint 1-71.
\bibitem {dk2} P.Deift, T.Kriecherbauer, K.T-R McLaughlin,
S.Venakides, X.Zhou, Uniform asymptotics for polynomials
orthogonal with respect to varying exponential weights and
applications to universality questions in random matrix theory,
preprint 1-106.
\bibitem {dem} K.Demeterfi, Two-dimensional quantum gravity, matrix
models and string theory, Int.J. Mod.Phys. {\bf A8} (1993) 1185-1244.
\bibitem {DiF}  P.Di Francesco, P.Ginsparg, J.Zinn-Justin,
2$D$ gravity and random matrices, Phys.Rep. {\bf 254} (1995) 1-133.
\bibitem {di2} P.Di Francesco, O.Golinelli, E.Guitter, Meander, folding and arch
statistics, in Combinatorics and Physics, Mathematical and Computer Modelling
144 (1996).
\bibitem {dup} B.Duplantier, I.K.Kostov, Geometrical critical phenomena on a
random surface of arbitrary genus, Nucl.Phys. {\bf B340} (1990) 491.
\bibitem {dur} B.Durhuus, C.Kristjansen, Phase structure of the
 $O(n)$ model on a random lattice for $n>2$,  Nucl.Phys. {\bf B483} (1997) 535-551.
\bibitem {ey1} B.Eynard, C.Kristjansen, Exact solution of the $O(n)$ model on a
random lattice, Nucl.Phys. {\bf B455} (1995) 577-618.
\bibitem {ey2} B.Eynard, C.Kristjansen, More on the exact solution of the $O(n)$
 model on a random lattice and an investigation of the case $|n|>2 $ ,
Nucl.Phys. {\bf B466} (1996) 463.
\bibitem {ek} B.Eynard, C.Kristjansen, An iterative solution of the three-colour
problem on a random lattice, Nucl.Phys. {\bf B516} (1998) 529.
\bibitem {fz} J.Feinberg, A.Zee, Renormalizing rectangles and other topics in
random matrix theory, cond-mat/9609190.
\bibitem {fz2} J.Feinberg, A.Zee, Non-gaussian non-hermitian random matrix
theory: phase transitions and addition formalism, Nucl.Phys. {\bf B501} (1997) 643.
\bibitem {fz3} J.Feinberg, A.Zee, Non-hermitian random matrix theory:
method of hermitian reduction, Nucl.Phys. {\bf B504} (1997) 579-608.
\bibitem {fi1} A.S.Fokas, A.R.Its andA.V.Kitaev, The isomonodromy
approach to matrix models in 2D quantum gravity, Comm.Math.Phys.
{\bf 147} (1992) 395-430.
\bibitem {fi2} A.S.Fokas, A.R.Its andA.V.Kitaev, Discrete Painleve
equations and their appearance in  quantum gravity,
Comm.Math.Phys. {\bf 142} (1991) 313-344.
\bibitem {fo} P.J.Forrester, The spectrum of random matrix ensembles, Nucl.Phys.
{\bf B402} (1993) 709.
\bibitem {fm} Y.V.Fyodorov, A.D.Mirlin, Statistical properties of eigenfunctions
 of random quasi 1D one particle hamiltonians, Int.J.Mod.Phys.  
{\bf B 8} (1994) 3795-3842.  
\bibitem {gau} M.Gaudin, I.Kostov,  $O(n)$ model on a fluctuating planar
lattice: some exact results, Phys.Lett.{\bf B220} (1989) 200.
\bibitem {gin} J.Ginibre, Statistical ensembles of complex, quaternion and real
matrices, Jour.Math.Phys. {\bf 6} (1965) 440-449.
\bibitem {gw} D.J.Gross, E.Witten, Possible third-order phase transition in the large-N
lattice gauge theory, Phys.Rev. {\bf D21} (1980) 446-453.
\bibitem {gro} D.J.Gross, T.Piran, S.Weinberg, Two dimensional quantum gravity
and random surfaces, World Scientific  (1992)
\bibitem {Guh}   T.Guhr, A.Muller-Groeling, M.A.Weidenmuller, Random matrix
theories in quantum physics : common concepts ,Phys.Rep. {\bf 299} (1998)
190.
\bibitem {hel} S.Helgason , Differential geometry, Lie groups and symmetric
spaces, Academic Press (1978).
\bibitem {hua} L.K.Hua, Harmonic analysis of functions of several complex
variables in the classical domains, AMS Providence, Rhode Island (1963).
\bibitem {ilg} E.M.Ilgenfritz, Yu.M.Makeenko, T.V.Shahbazyan, On the relation between many-color QCD and the free Nambu string on the lattice,
Phys. Lett. {\bf B172} (1986) 81-85.
\bibitem {IK}  A.R.Its and A.V.Kitaev, Mathematical aspects of the non-perturbative 2$D$
 quantum gravity, Mod.Phys.Lett.A5 (1990) 2079.
\bibitem {pol1} R.A.Janik, M.A.Nowak, G.Papp, I.Zahed, Nonhermitian random
matrix models,  Nucl.Phys. {\bf B501} (1997) 603.
\bibitem {pol2}  R.A.Janik, M.A.Nowak, G.Papp, I.Zahed,  The U(1) problem in chiral random matrix models, Nucl.Phys. {\bf B498} (1997) 313-330.
\bibitem {pol5} R.A.Janik, M.A.Nowak, G.Papp, I.Zahed,  Various shades of Blue's
functions, 1997 Zakopane lectures, hep-th/9710103.
\bibitem {jur} J.Jurkiewicz, K.Zalewski, Phase structure of $U(N \to \infty)$ gauge
theory on a two-dimensional lattice for a broad class of variant actions,
 Nucl.Phys. {\bf B220} (1983) 167-184.
\bibitem {ju1} J.Jurkiewicz, Regularization of one-matrix models, Phys. Lett.
{\bf B245} (1990) 178-184.
\bibitem {ju2} J.Jurkiewicz, Chaotic behaviour in one-matrix models, Phys. Lett.
{\bf B261} (1991) 260-268.
\bibitem {kf} E.Kanzieper, V.Freilikher, Spectra of large random matrices: a
method of study, (1998) cond-mat/9809365.
\bibitem {ka1} V.A.Kazakov, Ising model on a dynamical planar random lattice:
exact solution , Phys.Lett.{\bf A119} (1986) 140-144.
\bibitem {ka8} V.A.Kazakov, Nucl.Phys.(Proc.Suppl.) {\bf B4} (1988) 93.
\bibitem {ka3} V.A.Kazakov, A simple solvable model of quantum field theory of
open strings,  Phys.Lett.{\bf B237} (1990) 212.
\bibitem {ka4} V.A.Kazakov, P.Zinn-Justin, Two matrix model with $ABAB$
interaction, Nucl.Phys. {\bf B546} (1999) 647 .
\bibitem  {ko} I.Kostov, $O(N)$ vector model on a planar random lattice :
spectrum of anomalous dimensions , Mod.Phys.Lett.{\bf A4} (1989) 217-226.
\bibitem  {kos} I.Kostov, M.Staudacher , Multicritical phases of the $O(n)$ model
on a random lattice,  Nucl.Phys. {\bf B384} (1992) 459.
\bibitem {Le1} O.Lechtenfeld, On eigenvalue tunneling in matrix models,
Int.J.Mod.Phys. {bf A7} (1992) 2335.
\bibitem {le1} O.Lechtenfeld, R.Ray, A.Ray, Phase diagram and orthogonal
polynomials in multiple well matrix models, Int.J.Mod.Phys. {\bf A6} (1991)
4491-4515.
\bibitem {le2} O.Lechtenfeld, Semiclassical approach to finite N matrix models,
 Int.J.Mod.Phys. {\bf A7} (1992) 7097-7118.
\bibitem {lm} E.H.Lieb, D.C.Mattis, Mathematical physics in one dimension, Academic
 Press 1966.
\bibitem  {mak} Y.Makeenko, H.Win Pe, Supersymmetric matrix models and the
meander problem, (1996) hep-th/9601139.
\bibitem {Meh}  M.L.Mehta, Random Matrices, 2nd ed., Academic Press, 1991.

\bibitem {mol} L.Molinari, Phase structure of matrix models through orthogonal
polynomials, J.Phys. {\bf A21} (1988) 1-6.
\bibitem {mol2}  L.Molinari, E.Montaldi , The large N magnetization in matrix
models revisited, Nuovo Cimento {\bf D15} (1993) 293.
\bibitem {mul} M.Mulase, Lectures on the asymptotic expansion of a hermitian
matrix integral (1998) math-ph/9811023.

\bibitem {vip2} R.C.Myers, V.Periwal, From polymers to quantum gravity: triple
scaling in rectangular random matrix models, Nucl.Phys. {\bf B390} (1993) 716-746.

\bibitem {na1} T.Nagao, K.Slevin, Nonuniversal correlations for random matrix
ensembles, J.Math.Phys. {\bf 34} (1993) 2075 ; Laguerre ensembles of random
matrices :  nonuniversal correlation functions, J.Math.Phys. {\bf 34} (1993)
2317.
\bibitem {na2} T.Nagao, P.J.Forrester, Asymptotic correlations at the spectrum
edge of random matrices, Nucl.Phys. {\bf B435} (1995) 401.
\bibitem {no1} M.A.Nowak, J.J.Verbaarschot , I.Zahed, Chiral fermions in the instanton
vacuum at finite temperature, Nucl.Phys. {\bf B325} (1989) 581-592.
\bibitem {pss} L.A.Pastur, M.Shcherbina, Universality of the local eigenvalue
statistics for a class of unitary invariant random matrix ensembles, J.Stat.Phys.
{\bf 86} (1997) 109.
\bibitem {ple} J.C.Plefka, Iterative solution of the supereigenvalue model,
 Nucl.Phys. {\bf B444} (1995) 333-352 ; The supereigenvalue model in the 
double-scaling limit, {\bf B448} (1995) 355-372.
\bibitem {ros} P.Rossi, M.Campostrini, E.Vicari, The large-N expansion of
unitary-matrix models, Phys.Rep. {\bf 302} (1998) 143-209.
\bibitem {sas} M.Sasaki, H.Suzuki, Matrix realization of random surfaces,
 Phys.Rev {\bf D43} (1991) 4015-4028.
\bibitem {ss} G.W.Semenoff, R.J.Szabo, Fermionic matrix models, Int.J.Mod.Phys.
{\bf A12} (1997) 2135-2292.
\bibitem {sen} D.Senechal , Chaos in the hermitian one-matrix model,
 Int.J.Mod.Phys. {\bf A7} (1992) 1491.
\bibitem {shi} Y.Shimamune, On the phase structure of large $N$ matrix models
and gauge models, Phys.Lett.{\bf B108} (1982) 407.  The authors
of \cite{cmm} and  \cite {cmm2} were unaware of this letter.
\bibitem {sh} J.Shohat , A differential equation for orthogonal polynomials,
Duke Math.J. {\bf5} (1939) 401.
\bibitem {shu} E.Shuryak,  J.J.Verbaarschot, Random matrix theory and spectral sum rules for the Dirac operator in QCD, Nucl.Phys. {\bf A560} (1993) 306-320.
\bibitem {si} P.G.Silvestrov, Summing graphs for random band matrices, Phys.Rev. {\bf E 55}
(1997) 6419-6432.  
\bibitem {sim} Y.A.Simonov, Chiral-symmetry breaking in the disordered QCD vacuum, Phys.Rev. {\bf D43} (1991) 3534-3540.
\bibitem {tHo} The discovery by Gerard 't Hooft of the topological expansion,
Nucl.Phys. {\bf B72} (1974) 461 is probably the origin of the interest of
quantum  field theorists in random matrix theory.
\bibitem {tra}  C.A.Tracy, H.Widom , Introduction to random matrices,
in Springer lect. notes in physics 424, ed.G.F.Helminck , Geometric and 
quantum aspects of integrable systems (1993) 103.
\bibitem {ve1} J.J.Verbaarschot, I.Zahed, Spectral density of the QCD Dirac
operator near zero virtuality , Phys.Rev.Lett. {\bf 70} (1993)
3852 ; J.J.Verbaarschot, The spectrum of the QCD Dirac operator
and chiral random matrix theory : the threefold way,
Phys.Rev.Lett. {\bf 72} (1994) 2531 ; J.J.Verbaarschot, I.Zahed,
Random matrix theory and QCD, Phys.Rev.Lett. {\bf 73} (1994) 2288.
\bibitem {wit} E.Witten, Baryons in the 1/N expansion, Nucl.Phys. {\bf B160} (1979) 57-115.
\bibitem {pz} P.Zinn-Justin, The dilute Potts model on random surfaces,
cond-mat/9903385.
\bibitem {pz2} P.Zinn-Justin, The six-vertex model on random lattices,
Europhys.Lett. {\bf 49} (2000) 15-21.

\end{thebibliography}
\end{document}